\documentclass[iop]{emulateapj}

\usepackage{amsmath}
\usepackage{rotating}		
\usepackage{morefloats}
\usepackage{graphics}
\usepackage{epsfig}
\usepackage{natbib}
\shorttitle{Lyman $\alpha$ in CTTSs}
\shortauthors{Schindhelm et al.}

\begin{document}

\title{Lyman $\alpha$ dominance of the Classical T Tauri FUV Radiation Field}

\author{
Eric Schindhelm\altaffilmark{1}, Kevin France\altaffilmark{2}, Gregory J. Herczeg\altaffilmark{3},  Edwin Bergin\altaffilmark{4}, Hao Yang\altaffilmark{5}, Alexander Brown\altaffilmark{2}, Joanna M. Brown\altaffilmark{6}, Jeffrey L. Linsky\altaffilmark{7}, Jeff Valenti\altaffilmark{8}}

\altaffiltext{1}{Southwest Research Institute, 1050 Walnut St., Suite 300, Boulder, CO 80303, USA; eric@boulder.swri.edu}
\altaffiltext{2}{Center for Astrophysics and Space Astronomy, University of Colorado, 389 UCB, 
Boulder, CO 80309, USA}
\altaffiltext{3}{Kavli Institute for Astronomy and Astrophysics, Peking University, Beijing 100871, China}
\altaffiltext{4}{Department of Astronomy, University of Michigan, 830 Dennison Building, 500 Church Street, Ann Arbor, MI 48109, USA}
\altaffiltext{5}{Institute of Astrophysics, Central China Normal University, Wuhan, Hubei, 430079, China}
\altaffiltext{6}{Harvard-Smithsonian Center for Astrophysics, 60 Garden Street, MS-78, Cambridge, MA 02138, USA}
\altaffiltext{7}{JILA, University of Colorado and NIST, 440 UCB, Boulder, CO 80309, USA}
\altaffiltext{8}{Space Telescope Science Institute, 3700 San Martin Drive, Baltimore MD, 21218, USA}


\begin{abstract}
Far-ultraviolet (FUV) radiation plays an important role in determining chemical abundances in protoplanetary disks.  \ion{H}{1} Lyman $\alpha$  is suspected to be the dominant component of the FUV emission from Classical T Tauri Stars (CTTSs), but is difficult to measure directly due to circumstellar and interstellar \ion{H}{1} absorption. To better characterize the intrinsic Lyman $\alpha$ radiation, we present FUV spectra of 14 CTTSs taken with the \emph{Hubble Space Telescope} COS and STIS instruments.  H$_2$ fluorescence, commonly seen in the spectra of CTTSs, is excited by Lyman $\alpha$ photons, providing an indirect measure of the Lyman $\alpha$ flux incident upon the warm disk surface.  We use observed H$_2$ progression fluxes to reconstruct the CTTS Lyman $\alpha$ profiles.  The Lyman $\alpha$ flux correlates with total measured FUV flux, in agreement with an accretion-related source of FUV emission. With a geometry-independent analysis, we confirm that in accreting T Tauri systems Lyman $\alpha$ radiation dominates the  FUV flux ($\sim$1150 \AA\ - 1700 \AA).  In the systems surveyed this one line comprises 70 - 90 \% of the total FUV  flux.  
\end{abstract}


\clearpage

\section{Introduction}

One of the most influential factors in determining the local composition and physical state of the gas in protoplanetary disks is the ultraviolet (UV) radiation field. Strong UV radiation is produced from hot gas in the magnetically active atmospheres of the central star, and at the magnetospheric shock where disk material collides with the stellar atmosphere \citep{GSchmitt2008,Sacco2008}. UV photons have a profound influence on the gas heating \citep{Jonkheid2004,Nomura2007,Woitke2009} and disk gas chemistry \citep{Aikawa1999,Bethell2009,Fogel2011}. Surveys of the UV radiation field have been able to characterize the general spectral structure of the radiation \citep{Valenti2000,Yang2012}. \ion{H}{1} Lyman $\alpha$ (Ly$\alpha$) is a significant contributor to the UV radiation, comprising as much as 80\% of the far-UV (FUV) emission produced in the stellar atmosphere and accretion shock \citep{Bergin2003,Herczeg2004}. While the Ly$\alpha$ radiation is intense close to the star, atomic hydrogen in the disk atmosphere at larger radii isotropically scatters Ly$\alpha$ photons which provides greater penetration into the molecular disk.  This is quite different from UV photons at other wavelengths that are scattered solely by dust grains, leading to a Ly$\alpha$ dominated radiation field - if Ly$\alpha$ radiation is present  \citep{Herczeg2005b,Bethell2011b}.  This is important because Ly$\alpha$ dissociates many molecules such as H$_2$O and CH$_4$, and can lead to an overabundance of certain species, especially CN \citep{Bergin2003,VanZadelhoff2003,VanDishoeck2006}.

While many strong UV transitions such as \ion{C}{4} and \ion{He}{2}  have been studied \citep{Ingleby2011b,Yang2012}, an intrinsic Ly$\alpha$ profile has only been well characterized in a single Classical T Tauri Star (CTTS) - TW Hya \citep{Herczeg2004}. Due to circumstellar and interstellar \ion{H}{1} absorption, it is impossible to directly measure the intrinsic Ly$\alpha$ emission of most CTTSs. On the other hand, H$_2$ emission resulting from  Ly$\alpha$ photoexcitation is prevalent throughout the FUV bandpass, providing an indirect probe of the Ly$\alpha$.  Several Lyman and Werner band transitions reside at wavelengths coincident with the Ly$\alpha$ profile, and have a relatively simple radiative transfer. These lines have been identified in previous observations of CTTSs \citep{Brown1981,Valenti2000,Ardila2002,Herczeg2002,Herczeg2006,France2011b}. H$_2$ emission therefore provides an opportunity to study the circumstellar region where the Ly$\alpha$ radiation interacts with the disk molecular gas, as well as characterize the strength of this key component of the FUV radiation field in T Tauri systems.

The impact of strong Ly$\alpha$ radiation on disks has only recently been explored in detailed UV radiation transfer models \citep{Bethell2011b} and chemical models \citep{Fogel2011}.  Given the influence of these photons on gas throughout the disk, quantifying the Ly$\alpha$ emission in CTTSs is crucial to produce accurate disk thermal/chemistry models.  In this letter, we reconstruct the Ly$\alpha$ profile from observed H$_2$ emission lines in the FUV spectra of CTTSs.

\section{Observations}
\label{observations}

Our sample includes 14 CTTSs, chosen for their strong H$_2$ and FUV continuum emission.  The targets, as well as their distances and extinctions, are listed in Table \ref{Target_Parameters}.  We adopt  target and line-of-sight parameters from \citet{France2012}.  Most observations utilized the G130M and G160M medium resolution ($\Delta v \approx$17 km s$^{-1}$) modes of COS \citep{Green2012}.  We also use archival STIS E140M observations of TW Hya from the StarCAT spectral catalog \citep{Ayres2010} to compare our results with previous analysis \citep{Herczeg2004}.  

This sample includes targets from the Taurus-Auriga, $\eta$ Chamaeleontis, and TW Hya star-forming regions, in addition to isolated systems. They are a subset of a full CTTS H$_2$ survey presented in more detail in France et al. (2012). All the CTTSs in our sample display Ly$\alpha$-excited H$_2$ emission lines throughout the COS bandpass with an average line width of 46 $\pm$ 14 km s$^{-1}$.  Circumstellar and interstellar \ion{H}{1} absorbs the Ly$\alpha$ profiles to varying degrees in each target.  Figure \ref{fig:Lyman_method} shows the observed H$_2$ emission and Ly$\alpha$ profiles of several targets.  An \ion{H}{1}  component absorbs the Ly$\alpha$ line core in V4046 Sgr, while an outflow absorber attenuates most of the short wavelength side of the Ly$\alpha$ profile of RU Lupi. Very little Ly$\alpha$ emission is observed in the DE Tau spectra.  However, each target shows abundant Ly$\alpha$ excited H$_2$ emission lines \citep{France2012}, demonstrating that the observed Ly$\alpha$ profile can vastly underrepresent the local Ly$\alpha$ radiation field at the disk surface.

\begin{table*}
{\bf \caption{\label{Target_Parameters} Target Properties and Fluxes}}\begin{center}
\begin{tabular}{ccccccccc}\hline
Target &  $A_V$ & d  & $<F_{Ly\alpha}>^a$ & $<F_{ab}>^b$ & FWHM  & $N_{out}$  & $v_{out}$ & $F_{FUV}$$^d$  \\ 
 & & (pc) & &  & (km s$^{-1}$)$^c$  & (10$^{19}$ cm$^{-2}$)$^c$  &  (km s$^{-1}$)$^c$  & \\

\hline
AA Tau &        0.5 &      140   & 35.2 $\pm$        6.3  & 16.5 $\pm$        2.4  &       642 $\pm$        33 &      1.11 $\pm$        0.31 &   -143 $\pm$       16 &  3.5\\
BP Tau &        0.5 &     140   &   31.5 $\pm$        5.5  & 20.7 $\pm$        3.0  &       613 $\pm$        54 &      0.44 $\pm$        0.30 &   -131 $\pm$        29 & 14.1\\
DE Tau &        0.6 &      140  &  15.4 $\pm$        4.5   & 7.91 $\pm$        2.8  &        623 $\pm$        75 &       0.92 $\pm$        0.53 &   -156 $\pm$        28 &   3.6\\
DM Tau &        0.0 &      140     &  4.6 $\pm$        0.5  & 3.19 $\pm$        0.4  &        912 $\pm$        35 &      0.39 $\pm$        0.03 &   -89 $\pm$        4 &     1.1\\
GM Aur &        0.1 &      140   &   12.2 $\pm$        3.0   & 6.19 $\pm$        1.5  &        815 $\pm$        31 &       1.16 $\pm$        0.32 &   -129 $\pm$        17 & 2.9\\
HN Tau &        0.5 &      140    &   12.9 $\pm$        2.6   & 7.08 $\pm$        1.4  &        776 $\pm$        40 &      1.15 $\pm$        0.49 &   -153 $\pm$        26 &  2.8\\
LkCa 15 &        0.6 &     140    &   17.2 $\pm$        2.8  & 9.54 $\pm$        2.0  &        665 $\pm$        39 &      0.80 $\pm$        0.40 &   -139 $\pm$        25 &  3.0\\
RECX 11 &        0.0 &      97   &   5.2 $\pm$        0.4   & 3.07 $\pm$        0.6  &        573 $\pm$        63 &     0.78 $\pm$        0.48 &   -126 $\pm$        35 &   0.8\\
RECX 15 &        0.0 &      97   &  11.6 $\pm$        1.6   & 5.73 $\pm$        1.0  &        848 $\pm$        54 &      0.96 $\pm$        0.38 &   -125 $\pm$        27 &  1.0\\
RU Lupi &        0.1 &      150  &  23.4 $\pm$        5.5   & 10.3 $\pm$        1.7  &       707 $\pm$        59 &      3.14 $\pm$        1.56 &   -219 $\pm$        52 &  7.5\\
SU Aur &        0.9 &   140    &   35.4 $\pm$        8.2    & 14.8 $\pm$        1.9   &       644 $\pm$        59 &      2.34 $\pm$        1.37 &   -177 $\pm$        47 &  6.9\\
TW Hya &        0.0 &    56   &   52.5 $\pm$        9.1   & 38.3 $\pm$        6.6   &       865 $\pm$        35 &     0.29 $\pm$        0.12 &   -77 $\pm$        16 &  41.4\\
UX Tau &        0.2 &     140   &   6.2 $\pm$        0.5   & 4.32 $\pm$        0.4  &        606 $\pm$        42 &     0.18 $\pm$        0.04 &   -64 $\pm$        7 &  1.1\\
V4046 Sgr &        0.0 &     83    &   46.0 $\pm$        5.1 & 24.9 $\pm$        2.2  &       627 $\pm$        38 &        0.65 $\pm$        0.26 &   -108 $\pm$        17 &  7.0\\
  \hline
\end{tabular}
\end{center}

Fluxes in 10$^{-12}$ ergs cm$^{-2}$ s$^{-1}$

$^a$ Intrinsic, unabsorbed model flux

$^b$ H$_2$-incident, circumstellar \ion{H}{1}-absorbed model flux

$^c$ Average parameters of intrinsic model Ly$\alpha$ profiles

$^d$ Integrated 1160 - 1695 \AA\ flux, excluding Ly$\alpha$ (1207 - 1222 \AA) and \ion{O}{1} (1300 - 1310 \AA).

\end{table*}

\section{Analysis}
\label{analysis}

\citet{Wood2002} and \citet{Herczeg2004} developed a method of Ly$\alpha$ reconstruction using photo-excited H$_2$ emission lines in the UV spectra of Mira B and TW Hya.  The line fluxes were used in fluorescence models to determine the intrinsic Ly$\alpha$ profiles and H$_2$ properties of each object. To make an analysis of a larger sample tractable, we employ a similar technique with only the brightest H$_2$ progressions. A progression refers to the cascade of electronic transitions originating from an individual rotational-vibrational level of the excited electronic state.  Similar to the previous studies, we first measure the H$_2$ lines, and then model the Ly$\alpha$ profiles.

\subsection{H$_{2}$ Emission Spectra and the Absorbed Ly$\alpha$ Flux}

We employed a multi-Gaussian IDL line-fitting code, optimized for COS emission line spectra, to measure the total flux from Ly$\alpha$-pumped H$_{2}$. This code assumes a Gaussian line-shape convolved with a wavelength dependent line spread function (LSF), then uses the MPFIT routine to minimize $\chi^{2}$ between the fit and data \citep{Markwardt2009}.  An unconvolved Gaussian was used for TW Hya observed with STIS.  To reconstruct the local Ly$\alpha$ profile incident upon the molecular disk surface, we measured the total flux from 12 fluorescent progressions excited by Ly$\alpha$.  We chose to restrict the emission line analysis to the 1395~--~1640~\AA\ range to mitigate the effects of H$_{2}$ self-absorption, which are strongest at $\lambda$~$\lesssim$~1400~\AA~\citep{Herczeg2004}.  The brightest, unblended lines from 12 progressions in the 1395~--~1640~\AA\ bandpass (up to 38 lines) were fitted. We refer the reader to France et al. (2012) for additional information on the H$_{2}$ emission characteristics of our sample, including progression IDs for all lines measured.

The total flux from progression $m$ is given by 
\begin{equation}
F_{m} = \frac{1}{N} \sum \left ( \frac{F_{mn}}{B_{mn}} \right ) 
\end{equation}
where $F_{mn}$ is the reddening corrected, integrated H$_{2}$ emission line flux (in units of ergs cm$^{-2}$ s$^{-1}$) from rovibrational state $m$ (= [$v^{'}$,$J^{'}$]) in the  $B$$^{1}\Sigma^{+}_{u}$ electronic state to $n$ (= [$v^{''}$,$J^{''}$]) in the ground electronic state, $X$$^{1}\Sigma^{+}_{g}$. $B_{mn}$ is the branching ratio for each transition in a progression, defined as the ratio of the Einstein $A$-value for a given transition $m$~$\rightarrow$~$n$ ($A_{v'J' \rightarrow v''J''}$) to the total transition rate out of state $m$, including transitions to bound states and the vibrational continuum~\citep{stecher1967,wood2004,France2011a}.   $N$ is the number of emission lines measured from a given progression.

We take the error to be the standard deviation of the individual measurements of $F_{m}$(H$_{2}$), as this is the relative uncertainty in H$_{2}$ fluxes in most cases.  These standard deviations are typically less than 10\% for bright progressions, and as high as 20 - 30\% in weaker progressions.  The primary uncertainty in H$_2$ luminosities relates to the extinction correction (see $A_V$ in Table 1), although to zeroeth order this would affect all progressions similarly as they have been chosen from a narrow range of wavelengths.  Upper limits on the H$_{2}$ emission line fluxes of undetectable progressions were determined from the standard deviation in a $\pm$~50 km s$^{-1}$ region surrounding the laboratory wavelength of the transition. 

The flux incident on the H$_2$ ($I_{inc}(\lambda_m)$, in units of ergs cm$^{-2}$ s$^{-1}$ \AA$^{-1}$) at each H$_{2}$ absorbing transition wavelength $\lambda_m$ was then assumed to be the total progression flux  divided by the equivalent width ($W_{\lambda}$) of the absorbing H$_{2}$ transition:

\begin{equation}
I_{inc}(\lambda_m) = F_{m}/W_\lambda
\end{equation}

\begin{equation}
W_\lambda= \int_{\lambda}\big(1-e^{-\tau_{\lambda}[T_{H_2},N_{H_2}]}\big) d\lambda
\end{equation}

\noindent where each [$v$,$J$] state of the ground electronic level is populated by a single rotational temperature ($T_{H_2}$) and column density ($N_{H_2}$).  Non-thermal populations can result from dust heating and photo excitation by the strong X-ray and UV radiation fields \citep{Nomura2007,Gorti2009}. However, a degeneracy exists between $T_{H_2}$ and $N_{H_2}$ in our model, discussed further in Section \ref{LyA_analysis}. As a first order approximation to simplify this degeneracy we assume a thermal population. To determine $W_{\lambda}$, we used a grid of $T_{H_2}$ in 100 K increments from 1000 K to 5000 K, and $N_{H_2}$ in 0.1 dex increments from 10$^{16}$ to 10$^{22}$ cm$^{-2}$.  Models with $T_{H_2}$ $<$ 1000 K produce a negligible population in the $v$=2 level of the ground electronic state and are therefore not considered. For each of the 12 transitions, we calculated a set of $I_{inc}$ for each pair of $T_{H_2}$ and $N_{H_2}$.  Each pair of $T_{H_2}$ and $N_{H_2}$ therefore creates 12 unique values of $I_{inc}$.

  \begin{figure*}[htp]
  \centering
  \includegraphics[width=13cm,angle=90]{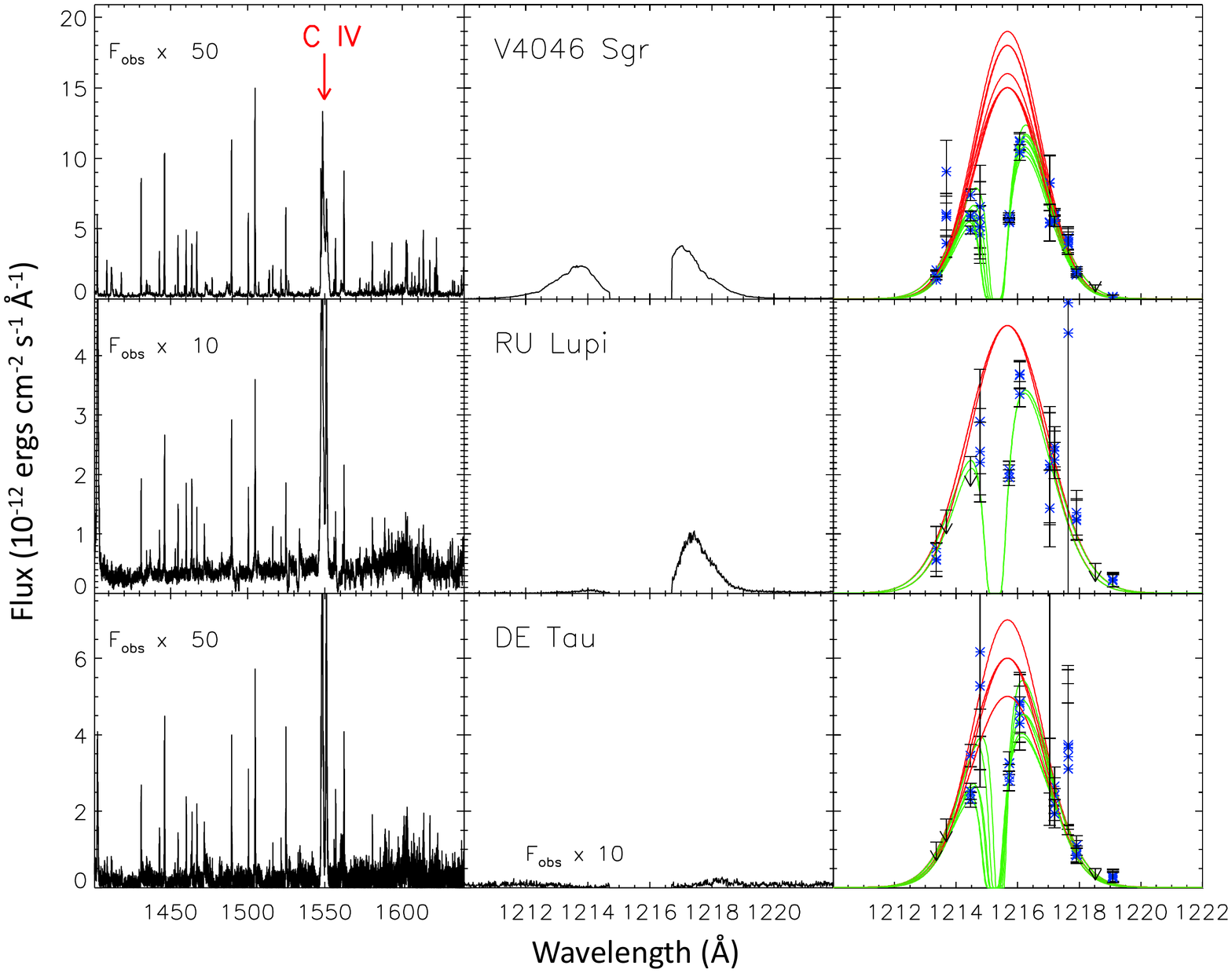}
  \caption{Observed and model spectra for V4046 Sgr (top row), RU Lupi (middle row), and DE Tau (bottom row).  H$_2$ emission lines are present throughout the FUV (left column), regardless of whether or not the photo-exciting Ly$\alpha$ is observed (middle).  In RU Lupi, a circumstellar wind extincts most of the short wavelength side of the Ly$\alpha$ profile, while in DE Tau Ly$\alpha$ is completely absorbed. We use the photo-excited H$_2$ lines to reconstruct model Ly$\alpha$ profiles (right).  Each pair of intrinsic ($I_{Ly\alpha}$, red) and outflow-absorbed ($I_{ab}$, green) profiles are a best fit to a different set of [$N_{H_2},T_{H_2}$]-based incident fluxes ($I_{inc}$, blue asterisks).}
   \label{fig:Lyman_method}
  \end{figure*}

\subsection{Ly$\alpha$ Profile Reconstruction}
\label{LyA_analysis}

Our reconstruction technique finds the set of outflow-absorbed Ly$\alpha$ profiles  ($I_{ab}$, in units of ergs cm$^{-2}$ s$^{-1}$ \AA$^{-1}$) that best fit the variety of $I_{inc}$ values.  The model assumes Ly$\alpha$ emission is created near the stellar surface.  Some amount of \ion{H}{1} lies in between the Ly$\alpha$ source and the H$_2$, absorbing some of the Ly$\alpha$ photons before they reach the H$_2$.  For each target, we create a grid of $I_{ab}$ profiles, each with its own single Gaussian emission component and single outflow \ion{H}{1} absorber: 


\begin{equation}
I_{ab} = (C_{Ly\alpha}+I_{Ly\alpha})e^{-\tau_{out}}
\end{equation}

 $C_{Ly\alpha}$ is the average continuum flux per Angstrom near Ly$\alpha$. $I_{Ly\alpha}$ is the accretion/magnetospheric-generated Gaussian emission profile, centered at the stellar radial velocity and parameterized by amplitude $I_0$ and FWHM (in units of ergs cm$^{-2}$ s$^{-1}$ \AA$^{-1}$ and km s$^{-1}$, respectively).  The optical depth $\tau_{out}$ is determined with a Voigt profile for the \ion{H}{1} Ly$\alpha$ transition, characterized by an outflow velocity $v_{out}$ and column density $N_{out}$ (in units of km s$^{-1}$ and cm$^{-2}$, respectively). This outflow absorption component is essential to adequately fitting the $I_{inc}$ values.

For each $I_{ab}$ model profile,  a $\chi^2$ is calculated with each set of [$T_{H_2}$,$N_{H_2}$]-dependent $I_{inc}$.  We then exclude any sets of $T_{H_2}$ and $N_{H_2}$ which yield a $\chi^2$ greater than 95.4\% (2$\sigma$) from the peak of the $\chi^2$ probability distribution.  The H$_2$ progressions for which only upper limits are calculated provide additional constraints on the Ly$\alpha$ profile;  we reject any Ly$\alpha$ profile with a flux greater than an upper limit value at that wavelength.  Given the various combinations of $I_0$, FWHM, $v_{out}$, $N_{out}$, $T_{H_2}$, and $N_{H_2}$, a variety of Ly$\alpha$ profiles adequately fit the $I_{inc}$ values.  This is demonstrated in Figure \ref{fig:Lyman_method}, where several combinations of $I_{Ly\alpha}$, $I_{ab}$, and $I_{inc}$ are shown.  We quantify the overall distribution of $I_{Ly\alpha}$  profiles for each target by integrating each $I_{Ly\alpha}$ profile from 1210 \AA\ to 1222 \AA, resulting in fluxes $F_{Ly\alpha}$  (in units of ergs cm$^{-2}$ s$^{-1}$) which form an approximately Gaussian distribution.  

\begin{figure*}
  \centering
  \includegraphics[width=16cm]{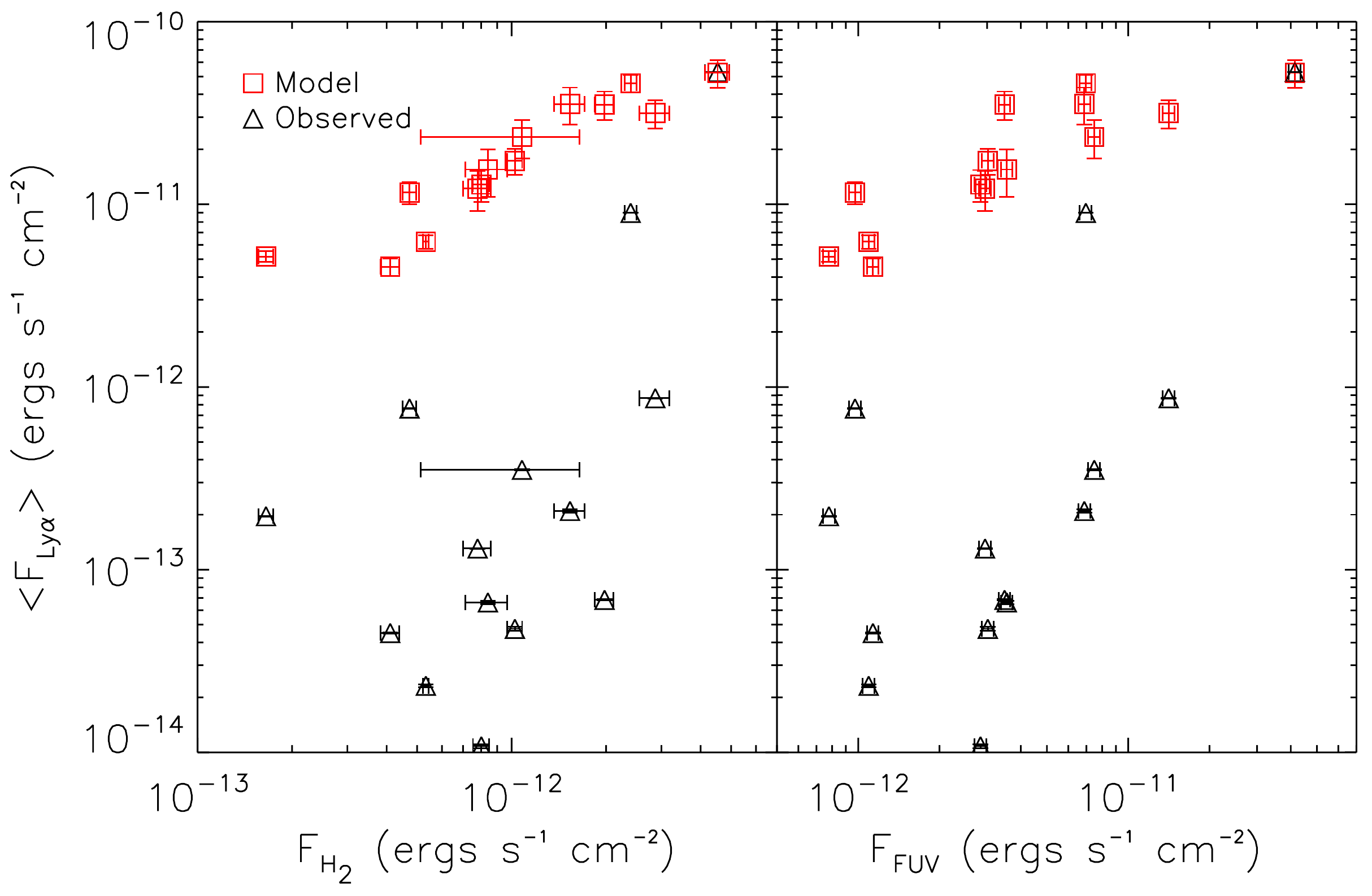}
  \caption{Integrated model (squares) and observed (triangles) Ly$\alpha$ flux vs. total H$_2$ flux (left) and total FUV flux (right) for each target.  The total H$_2$ flux is calculated by summing all the progression fluxes measured in our analysis.  $F_{FUV}$ is the total unreddened FUV flux measured over the 1160~--~1695~\AA\ bandpass.}
   \label{fig:Lyman_Fluxes}
  \end{figure*}

The average $F_{Ly\alpha}$ ($<F_{Ly\alpha}>$) of each target is listed in Table \ref{Target_Parameters}, along with the 1$\sigma$ width of the distribution. $<F_{Ly\alpha}>$ represents the total Ly$\alpha$ flux at the star, and is generally constrained to within 10 - 20 \%.  We also include the average emission FWHM and \ion{H}{1} absorber properties ($N_{out}$ and $v_{out})$.   Integrating $I_{ab}$, the H$_2$-incident Ly$\alpha$ flux, yields $<F_{ab}>$. Assuming a thermally populated ground state distribution, typical $T_{H_2}$ values are $\sim$2500 $\pm$ 1000 K and log($N_{H_2}$(cm$^{-2}$)) $\sim$19 $\pm$ 1, similar to those found in TW Hya \citep{Herczeg2004}. Strong UV and X-ray irradiation from the central star would act to preferentially populate excited rovibrational levels in excess of the thermal distribution (see e.g., \citet{Nomura2007}). In our simple model, non-thermal excitation would be approximated by higher rotational temperatures, and this increase in rovibrational population will be offset by lower total H$_{2}$ column densities.   The varying effects of non-thermal excitation most likely contribute to the spread of the observed $N_{H_2}$ and $T_{H_2}$ values.  Lower $N_{H_2}$ can require increased Ly$\alpha$ flux, although we expect that this increase will be of the same order of the uncertainty on the derived Ly$\alpha$ emission flux ($\sim$~10~--~20 \%).   Untangling the effects of non-thermal populations requires more complex modeling, including the effects of H$_{2}$ formation on dust grains, and disk modeling efforts that predict the UV fluorescent spectrum of H$_{2}$ (and CO) would be very valuable. Regardless, the overall model Ly$\alpha$ properties follow approximately gaussian distributions. Figure \ref{fig:Lyman_Fluxes} compares $<F_{Ly\alpha}>$ and the observed Ly$\alpha$ fluxes ($F_{obs}$, summed over the entire profile) with the total H$_2$ fluxes.  $F_{obs}$ has a Spearman $\rho$ rank correlation coefficient of 0.57 while $<F_{Ly\alpha}>$, by design, has a $\rho$ of 0.96.

Ideally we would compare our model Ly$\alpha$ profiles with the observed Ly$\alpha$ profiles.  However, this involves additional parameters and a more complex analysis than presented here.  Primarily, the radiative transfer of Ly$\alpha$ photons between the source, the H$_2$, and Earth must account for the geometric filling fraction $\eta$.  This fraction represents the solid angle of H$_2$ illuminated by the Ly$\alpha$ emission, such that $\eta$=1 implies the H$_2$ completely surrounds the star. An $\eta$ other than 1 would enter Equation 2 by multiplying $I_{inc}(\lambda_m)$ by $\eta$.  \citet{Herczeg2004} found $\eta$=0.25 for TW Hya, implying gas in a disk surface layer.  However, \citet{Wood2002} measured $\eta$ $>$ 2 in Mira B. A filling factor greater than 1 would possibly correspond to preferential scattering of H$_2$ photons into, or more likely Ly$\alpha$ photons out of, our line of sight \citep{Wood2002}. Future analysis will incorporate $\eta$ into a more detailed radiative transfer model of our sample targets.

\subsection{Total Ly$\alpha$ Flux in CTTSs}
\label{discussion}


We determine the fraction of the total FUV output from CTTSs that is in the Ly$\alpha$ emission line,  adopting a definition of the Ly$\alpha$ fraction, $f_{Ly\alpha}$, given by 
\begin{equation}
f_{Ly\alpha} =   \frac{F_{Ly\alpha}}{\left ( F_{FUV} + F_{Ly\alpha}  \right ) }
\label{fraction_equation}
\end{equation}

\noindent where  $F_{Ly\alpha}$ is our average model Ly$\alpha$ flux from the star (in units of ergs cm$^{-2}$ s$^{-1}$; \S\ref{LyA_analysis}) and $F_{FUV}$ is the total unreddened FUV flux, measured over the 1160~--~1695~\AA\ bandpass (in units of ergs cm$^{-2}$ s$^{-1}$).   The  wings of the stellar Ly$\alpha$ can extend to $>$~$\pm$~1000 km s$^{-1}$ from line center in some targets (see e.g, Figure 2 ($top$) from Yang et al. 2011), therefore we excluded the 1207~--~1225~\AA\ region from the computation of $F_{FUV}$.\nocite{Yang2011}  We also excluded the 1300~--~1310~\AA\ region to remove contamination by the geocoronal \ion{O}{1} multiplet.  A 5\% error is assumed on $F_{FUV}$ corresponding to the uncertainty in flux calibration.  The restriction of the total FUV band to 1160~--~1695~\AA\ is mainly driven by the the lack of supporting $FUSE$ observations at short wavelengths (912~--~1160~\AA) for all targets and the end of the STIS E140M bandpass at 1700~\AA.

Molecular line emission from H$_{2}$~(Herczeg et al. 2006; this work) and CO~\citep{France2011b,Schindhelm2012} contributes to $F_{FUV}$.  The FUV molecular emission is thought to originate from an inner warm disk surface ($a\le$ 3 AU) ~\citep{Herczeg2004,France2012} or outflow~\citep{Saucedo2003,Walter2003}.  The total FUV flux also includes a molecular continuum whose excitation mechanism and spatial distribution are not well constrained \citep{Bergin2004,France2011a}. $<F_{Ly\alpha}>$ shows a much better correlation with $F_{FUV}$ than $F_{obs}$  (Figure \ref{fig:Lyman_Fluxes}).  This is consistent with the idea that both the Ly$\alpha$ and FUV continuum emission are produced by related processes.

\begin{figure}
  \centering
  \includegraphics[width=9cm]{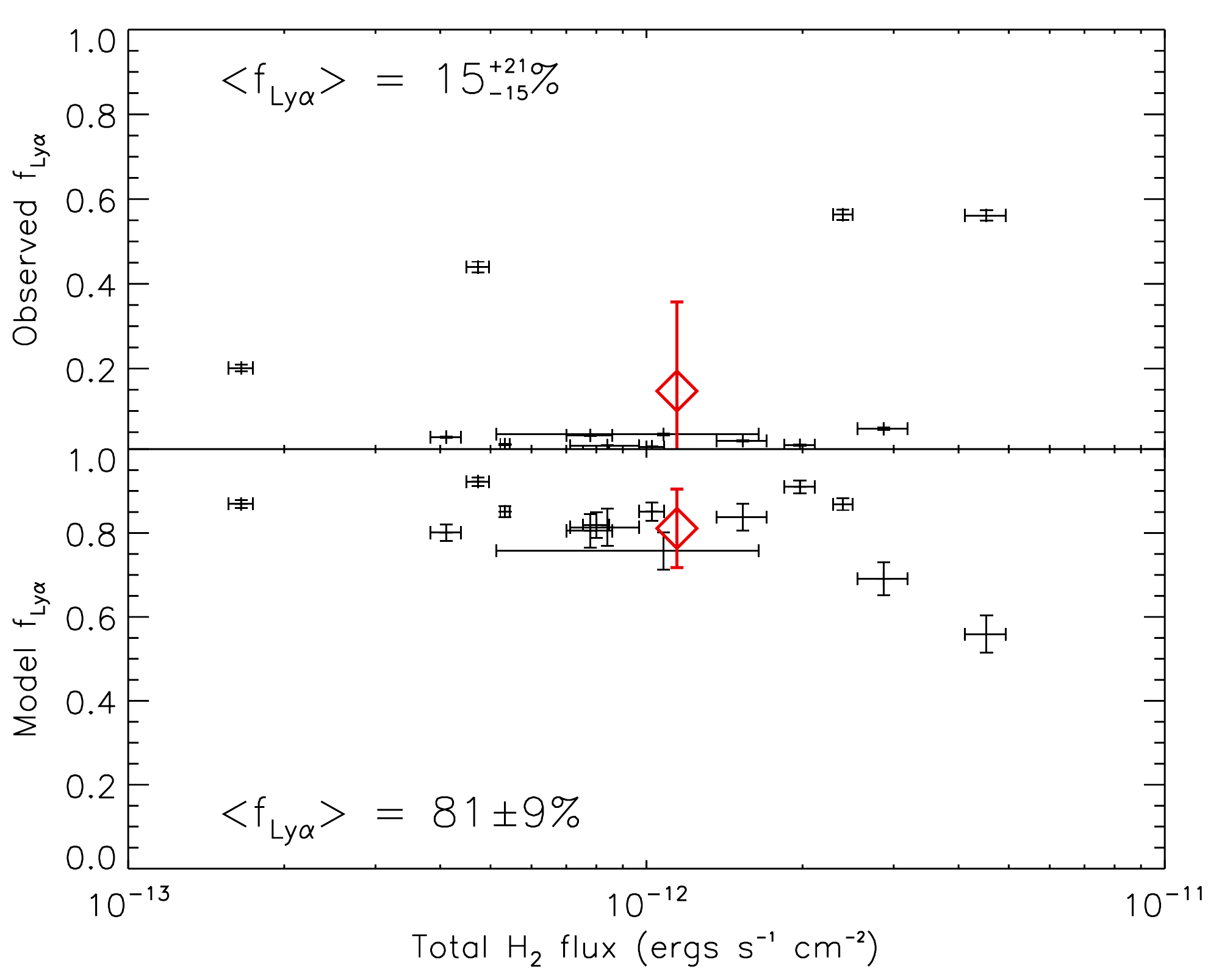}
  \caption{Observed (top) and model (bottom) Ly$\alpha$ fraction (assuming $\eta$=1) vs. total H$_2$ flux.  The diamond in each plot designates the average $f_{Ly\alpha}$ value for the observed and model profiles.  The significant increase in Ly$\alpha$ fraction from the observed to the model profiles demonstrates the importance of Ly$\alpha$ reconstruction.}
   \label{fig:Lyman_cont_frac}
  \end{figure}

We also calculate an observed Ly$\alpha$ fraction ($f_{obs}$) using $F_{obs}$ instead of $F_{Ly\alpha}$ in Equation \ref{fraction_equation}. We compare $f_{obs}$ and  $f_{Ly\alpha}$ with the total H$_2$ flux in Figure \ref{fig:Lyman_cont_frac}.  For our general assumption of $\eta$=1, we find an average $<f_{Ly\alpha}>$ of 81$\pm$9\%, compared with 15$^{+21}_{-15}$\% for $<f_{obs}>$.  For $\eta$ = 0.1, 0.25, 0.5, and 2.0, $<f_{Ly\alpha}>$ is 97\%, 94\%, 90\%, and 69\%, respectively.   A slight trend for decreasing $<f_{Ly\alpha}>$ with increasing $F_{FUV}$ is apparent, although this may be inappropriately weighted by BP Tau and TW Hya.  Regardless, our calculated $<f_{Ly\alpha}>$ values demonstrate the validity of previous assertions that Ly$\alpha$ emission dominates the FUV radiation field from CTTSs (e.g. \citet{Bergin2003}).

\section{Discussion and Conclusions}

We have demonstrated that Ly$\alpha$ dominates the FUV emission of CTTSs. Ly$\alpha$-fluoresced H$_2$ emission lines appear throughout the spectra of each CTTS in our survey. Circumstellar and interstellar \ion{H}{1} partially attenuates the line centers of the Ly$\alpha$ profiles in most targets and completely absorbs them in several targets.  The $observed$ Ly$\alpha$ fluxes do not correlate with either the total H$_2$ emission or the summed FUV continuum flux. The strongest fluorescent H$_2$ progressions are used to reconstruct the Ly$\alpha$ profile incident on the molecular disk, yielding $F_{Ly\alpha}$ values accurate to within $\sim20$\% in most targets.   This uncertainty is caused by the distribution of possible \ion{H}{1} and H$_2$ properties.  Our model assumptions (such as a single Gaussian emission component or  simple outflow absorber) may hide larger systematic errors, however future work will study these effects in a more detailed model.  We find that the intrinsic Ly$\alpha$ comprises 81$\pm9$\% of the total FUV emission from CTTSs, compared with a fraction of only 15$^{+21}_{-15}$\% for the observed Ly$\alpha$ profiles. This demonstrates the need for Ly$\alpha$ reconstruction to achieve accurate disk models of CTTSs.

It is clear from our results that the detection of the strong Ly$\alpha$ line in TW Hya was not limited to a single object.  Rather, strong Ly$\alpha$ emission dominates the FUV flux from all accreting (classical) T Tauri stars. Our measurements of the relative Ly$\alpha$/FUV continuum flux only compare Ly$\alpha$ to the FUV flux from 1160 - 1695 \AA.  Thus if there is significant UV flux shortward of 1160 \AA\ then the strength of Ly$\alpha$ relative to the FUV radiation could go down.  However, in TW Hya the flux below this limit is only 5\% of the FUV flux below 1700 \AA.  Similarly, \citet{France2011a} have shown that the FUV continuum decreases to shorter wavelengths across the FUV bandpass. In addition, the absorption properties of grains strongly limit the propagation of UV photons near the Lyman limit.  

The derived Ly$\alpha$ fractions confirm the dominance of Ly$\alpha$ in the FUV spectrum of the accreting young stars with disks.  This is important because Ly$\alpha$ photons from the star will see the atomic hydrogen layer, with shallow angle of incidence, above the molecular surface.  Isotropic scattering will lead a significant fraction of the Ly$\alpha$ flux to rain down on the disk with greater penetrating power than typical UV continuum photons \citep{Bethell2011b}.  In addition, the Ly$\alpha$ emission reprocessed by H$_2$ (which is scattered throughout the FUV spectrum) will also emit directly on the disk surface \citep{France2012}.  These two effects increase the penetration of UV photons beyond the simple case where one assumes the UV photons observe the disk surface with shallow angle of incidence and with the propagation solely influenced by grains. In general this should lead to greater heating and additional chemical effects deeper in the disk system.

ES and KF thank Brian Wood for input on Ly$\alpha$ profile reconstruction. This work was supported by NASA grants NNX08AC146 and NAS5-98043 to the University of Colorado at Boulder ($HST$ programs 11533 and 12036) and made use of data from $HST$ GO programs 8041 and 11616.







\end{document}